\pdfoutput=1
\documentclass[11pt]{article}
\usepackage{jheppub}

\usepackage{customprelude}
\usepackage{tikz}
\usetikzlibrary{positioning}

\title{\centering Ubiquity of non-geometry in heterotic compactifications}

\author[\clubsuit]{I\~naki Garc\'ia-Etxebarria,}
\author[\clubsuit,\heartsuit]{Dieter L\"ust,}
\author[\diamondsuit,\heartsuit]{Stefano Massai,}
\author[\heartsuit]{and Christoph Mayrhofer}

\affiliation[\clubsuit]{Max Planck Institute for Physics, F\"ohringer
  Ring 6, 80805 Munich, Germany}
\affiliation[\heartsuit]{Arnold Sommerfeld Center for Theoretical
  Physics, Theresienstra{\ss}e 37, 80333 M\"unchen, Germany}
\affiliation[\diamondsuit]{Enrico Fermi Institute, University of
  Chicago, 5640 S Ellis Ave, Chicago, IL 60637, USA}

\emailAdd{inaki@mpp.mpg.de}
\emailAdd{dieter.luest@lmu.de}
\emailAdd{massai@uchicago.edu}
\emailAdd{christoph.mayrhofer@lmu.de}

\abstract{We study the effect of quantum corrections on heterotic
  compactifications on elliptic fibrations away from the stable
  degeneration limit, elaborating on a recent observation by
  Malmendier and Morrison. We show that already for the simplest
  non-trivial elliptic fibration the effect is quite dramatic: the
  $I_1$ degeneration with trivial gauge background dynamically splits
  into two T-fects with monodromy around each T-fect being (conjugate
  to) T-duality along one of the legs of the $T^2$. This implies that
  almost every elliptic heterotic compactification becomes a
  non-geometric T-fold away from the stable degeneration limit. We
  also point out a subtlety due to this non-geometric splitting at
  finite fiber size. It arises when determining, via
  heterotic/F-theory duality, the SCFTs associated to a small number
  of pointlike instantons probing heterotic ADE singularities. Along
  the way we resolve various puzzles in the literature.}

\begin{document}


\makeatletter
\let\old@fpheader\@fpheader
\renewcommand{\@fpheader}{\old@fpheader\hfill
MPP-2016-333

\hfill LMU-ASC 60/16}
\makeatother

\maketitle

\section{Introduction}

In this note we study the behavior of the heterotic string on an
elliptically fibered K3, in the regime where the ten-dimensional gauge
group stays unbroken. In other words, the full non-triviality of the
gauge bundle is concentrated at points on the K3, where NS5 branes are
located \cite{Witten:1995gx}. While this is well-trodden territory, we
will show that one important aspect of the quantum dynamics of this
background has been overlooked.

The result is simplest to state, and most striking, in the case of the
local background given by the $I_1$ degeneration of the elliptic
fibration, in the absence of neighboring NS5 branes, i.e.\ a purely
geometric background. The total space of the elliptic fibration is in
this case smooth. We will take the size of the fiber to be finite. We
show that in this very simple background, quantum effects lead to a
splitting of the $I_1$ degeneration into a pair of non-geometric
T-fects---using nomenclature borrowed from
\cite{Lust:2015yia,Font:2016odl}. Around each of these defects of the
background, the monodromy on the $T^2$ fiber of the fibration is
(conjugate to) T-duality along one of the legs of the $T^2$. This
implies that almost every heterotic background is described by a
T-fold \cite{Hull:2004in}, once we take the quantum dynamics into
account.  The relevant quantum dynamics is in fact not exotic: in the
case of the $SO(32)$ heterotic string the relevant effects are
worldsheet-instantons wrapping the $T^2$ fiber, which dualize to the
well-known D$(-1)$-instanton effects in F-theory that split the O7$^-$
plane into its $B$ and $C$ components
\cite{Sen:1996vd,Sen:1997gv,Banks:1996nj}.

The physical reason for why the splitting happens is analogous to the
reason for the familiar O7$^-$ splitting: in the heterotic string we
have a modified Bianchi identity that gives rise to a ``wrong sign''
monodromy for the integral of the $B_2$-field over the $T^2$ fiber as
we go around an $I_1$ degeneration.  We have that, encircling $I_1$
counterclockwise, $b$ goes to $b-1$, in conventions where $b$ is the
real part of the complexified K\"ahler modulus
$\rho=\int_{T^2}B_2 +i \,{\rm Vol}(T^2)$ and the monodromy around an
NS5 is $b\to b+1$.  In order to obtain such a monodromy from the
na\"ive solution to the equation of motion for $\rho$ would force us
to set the volume of the $T^2$ fiber to negative values.\footnote{The
  argument here is analogous to the one leading to the na\"ive
  negative IIB string coupling $g_s$ close to an O7$^-$. For a review
  of the details see for instance \cite{Denef:2008wq} and references
  therein.} The heterotic string cures this sickness of the
innocent-looking $I_1$ degeneration by becoming a T-fold. In other
words, the ``bare'' heterotic $I_1$ splits into two T-fects with the
combined monodromy around the two T-fects being the desired
$b\to b-1$, but each T-fect having a more involved monodromy.

This viewpoint also suggests that once we put one or more NS5 branes
on top of the $I_1$ degeneration of the fibration there should be no
quantum splitting anymore, and one should recover the geometric
interpretation. This is indeed the case, as we will show.

\medskip

As in
\cite{McOrist:2010jw,Malmendier:2014uka,Font:2016odl,Malmendier:2016hji},
our approach to understand the non-geometry of the heterotic string
will be to use the map to F-theory \cite{Vafa:1996xn}. For our
purposes it will suffice to understand the duality map in the case of
unbroken gauge symmetry,\footnote{There has been recent interesting
  progress in understanding the $E_7\times E_8$ case
  \cite{Malmendier:2014uka,Gu:2014ova,Malmendier:2016hji}, and in fact
  our discussion sharpens the observation in \cite{Malmendier:2014uka}
  that the moduli space of the non-geometric heterotic $E_7\times E_8$
  on K3 agrees with the full moduli space of the family of F-theory
  geometries dual to certain geometric heterotic compactifications, as
  we discuss in more detail in \S\ref{sec:previous-work}.}  which was
worked out in the stable degeneration limit in
\cite{Morrison:1996na,Morrison:1996pp}, and away from the stable
degeneration limit (the case of most interest to us) in
\cite{LopesCardoso:1996hq} (see also
\cite{Lerche:1998nx,McOrist:2010jw,shiodaSandwich}). In
\S\ref{sec:I1-splitting} we will use this map for the simplest system
of an $I_1$ degeneration of an elliptic fibration over flat space,
with and without an NS5 brane. We will show that generically the $I_1$
degeneration splits, and we will analyze the nature of its T-fect
constituents. In \S\ref{sec:SO(32)} we analyze the case of the
$SO(32)$ heterotic string, and relate the quantum splitting in this
case to the well known splitting of the O7$^-$ plane in the dual
F-theory description.  In~\S\ref{sec:previous-work} we explain how our
observation sheds light on various obscure points in the
literature. We finish in~\S\ref{sec:discussion} by listing some
interesting open questions.

\section{Heterotic/F-theory duality map}

For simplicity, during most of our analysis we take a noncompact base
for the elliptic fibration, so we have an elliptic fibration over
$\bC$, which we parameterize by a complex coordinate $s$. We will
denote the coordinates on the elliptic fiber, given in Weierstra\ss{}
form, by $[x : y : z]$. The dual F-theory background is given
by a K3 fibration over our $\bC$ base, with the K3 itself elliptically
fibered. We denote by $[u : v]$ the coordinates on the $\bP^1$
base of the K3, and reuse $[x : y : z]$ for the coordinates of
the $\bP^{2,3,1}$ in which the elliptic fiber is embedded. It should
be always clear from the
context which $T^2$ we are discussing. In this section we will focus
on the case of the $E_8\times E_8$ heterotic string, leaving the
analysis of the $SO(32)$ heterotic to \S\ref{sec:SO(32)}.

For the convenience of the reader, let us first review the eight dimensional
map explicitly constructed in
\cite{LopesCardoso:1996hq,Lerche:1998nx,shiodaSandwich}. Start with an elliptically fibered K3 (giving the
fibration of the F-theory background) of the form
\begin{equation}
  \label{eq:K3}
  y^2 = x^3 + au^4v^4xz^4 + (bu^5v^7 + cu^6v^6 + du^7v^5)z^6\, .
\end{equation}
Parameterizing the complex structure and complexified K\"ahler modulus
of the dual heterotic $T^2$ by $\tau$ and $\rho=\int_{T^2}(B+iJ)$, respectively, we have
\begin{subequations}
  \label{eq:Cardoso-map}
  \begin{align}
    j(\tau)j(\rho) & = -12^6 \frac{a^3}{27bd}\, ,\\
    \lambda(\tau)\lambda(\rho) & = 12^6\frac{c^2}{4bd}
  \end{align}
\end{subequations}
where we have introduced $\lambda = 12^3-j$. In the holomorphic
setting\footnote{The duality group of the heterotic string on $T^2$ is
  $O(2,2;\bZ)$, which can be seen to be isomorphic to
  $(SL(2,\bZ)_\rho\times SL(2,\bZ)_\tau)\rtimes (\bZ_2\times \bZ_2)$,
  where the first $\bZ_2$ exchanges $\rho$ and $\tau$ (i.e.\ it is
  given by T-duality along one of the legs of the $T^2$) and the
  second $\bZ_2$ corresponds to a reflection along one of the
  directions of the $T^2$, which flips the choice of complex
  structure. We restrict to holomorphic fibrations, and thus ignore
  this second $\bZ_2$ factor. See \cite{Giveon:1994fu} for a review.}
$\tau$ and $\rho$ are only defined up to the duality action
$(SL(2,\bZ)_\rho\times SL(2,\bZ)_\tau)\rtimes \bZ_2$. Hence a map such as
\eqref{eq:Cardoso-map}, given by a symmetric function on the
$j$-functions, is the best that we can hope for.

Before proceeding, let us briefly recall some useful basic facts about
elliptic curves. It will be useful below to introduce
$\kappa=j/\lambda$. Given a particular $j$ invariant for an elliptic
curve, one can construct an elliptic curve with the same invariant by
taking the hypersurface
\begin{equation}
  \label{eq:kappa-elliptic-curve}
  y^2 = x^3 + 3\kappa x z^4 + 2\kappa z^6
\end{equation}
in $\bP^{2,3,1}$, as one can easily check. For any choice of the
coefficients $f$ and $g$, appearing in the Weierstra\ss{} representation, of the elliptic
curve with a given $j$-invariant, one has that
\begin{equation}
  \label{eq:lambda}
  \lambda = 12^3 - j = 12^3 \left(1 - \frac{4f^3}{4f^3+27g^2}\right) = 12^3
  \frac{27g^2}{4f^3 + 27g^2}
\end{equation}
and thus
\begin{equation}
  \label{eq:kappa-fg}
  \kappa=\frac{4f^3}{27g^2}\, .
\end{equation}
Notice that with these definitions the map~\eqref{eq:Cardoso-map}
implies
\begin{equation}
  \label{eq:kappa-Cardoso}
  \kappa(\rho)\kappa(\tau) = -\frac{4a^3}{27c^2}\, .
\end{equation}

Coming back to~\eqref{eq:Cardoso-map}, we can easily solve for
$j(\tau)$ and $j(\rho)$ in terms of the complex
structures of the K3 \eqref{eq:K3}. Introducing\footnote{We note that the quantity $\daleth$ is
  precisely the same (away from the location of the NS5 branes) as the
  quantity $q$ defined in \cite{Malmendier:2014uka} once we set $c=0$
  in the notation of that paper. We will explain this observation in
  \S\ref{sec:previous-work}.}
\begin{equation}
  \label{eq:daleth}
  \daleth = 12^3 a^3 bd + \left[4a^3 + 27(c^2 - 4bd)\right]^2\,,
\end{equation}
we have
\begin{equation}
  \label{eq:jtau-daleth}
  j(\tau) = -8\frac{4 \, a^{3} + 27 (c^{2} - 4 b d) -
            \sqrt{\daleth}}{bd}
\end{equation}
and
\begin{equation}
  \label{eq:jrho-daleth}
  j(\rho) = -8\frac{4 \, a^{3} + 27 (c^{2} - 4 b d) +
    \sqrt{\daleth}}{bd}\, .
\end{equation}
We have chosen a specific branch of the square root when writing these
formulas. Once we fiber the eight dimensional moduli over a complex
one-dimensional base, this choice of sign is the one agreeing with the
convention in which an NS5 induces a monodromy on $\rho$, leaving
$\tau$ invariant. Changing the sign corresponds to exchanging
$j(\tau)$ and $j(\rho)$, and is thus related to T-duality along one of
the legs of the $T^2$ fiber. Accordingly, we have that around single
zeroes of $\daleth$ we have a monodromy involving the T-duality
generator.  In other words, if $\daleth$ is not a perfect square (as a
function of $s$) the heterotic dual will be a T-fold.  As we will
show, this is precisely the case for a bare $I_1$ degeneration, i.e.\
one where the tadpole is not (locally) saturated.  Note that when
$\daleth$ is a perfect square, the heterotic moduli only have
monodromies in the
$SL(2,\mathbb{Z})_\rho \times SL(2,\mathbb{Z})_\tau$ subgroup of the
T-duality group. We emphasize that these backgrounds with reduced
monodromy can still be non-geometric if the monodromies do not admit
a classical interpretation globally, see for example
\cite{Hellerman:2002ax,McOrist:2010jw}.

\subsubsection*{Location of the NS5 branes}

In the stable degeneration limit we know that the NS5 branes are
located at the zeroes of $b$ and $d$
\cite{Morrison:1996na,Morrison:1996pp} (whether it is $b$ or $d$ that
vanishes depends on a tensor branch modulus, encoding the choice of
the $E_8$ factor associated with the small instanton). This statement
is also true away from the stable degeneration limit, as we now review.

Notice first that \eqref{eq:daleth} factorizes when $bd=0$:
\begin{equation}
  \label{eq:daleth-NS5}
  \daleth|_{bd=0} = \bigl[4a^3 +27c^2\bigr]^2\, .
\end{equation}
Furthermore, we see from~\eqref{eq:jtau-daleth} and \eqref{eq:jrho-daleth} that 
$j(\tau)$ stays finite at these points while $j(\rho)$ diverges. This is the
expected behavior for an NS5 brane. Hence, we find that the location of
the NS5 brane stays unmodified as we go away from the stable
degeneration limit.
Notice also that generically $\daleth|_{bd=0}\neq 0$, so an isolated
NS5 brane does not give rise to non-geometry, as expected from general
considerations.

\section{Quantum splitting of the $I_1$ degeneration in the 
  $E_8\times E_8$ heterotic string}
\label{sec:I1-splitting}

In the following, we turn to the details of the local, heterotic $E_8\times E_8$ compactification on $I_1$ with a locally trivial  gauge bundle. Already in this innocent looking case the above described T-fold behavior appears.
Following \cite{McOrist:2010jw,Font:2016odl}, we will study
this setup away from the large-volume limit for the fiber torus by
fibering adiabatically the eight dimensional heterotic/F-theory duality over a
common base, thus obtaining a six dimensional map.

\subsection{One NS5 on top of the $I_1$ degeneration}

We will start by considering the system composed of a small instanton
on top of an $I_1$ degeneration of the K3. For simplicity, we locate
the $I_1$-NS5 system at $s=0$. Due to the modified Bianchi identity
there is no monodromy for $\rho$ around $s=0$, while there is a
$\tau\to \tau+1$ monodromy. We expect this system to admit a geometric
description even after quantum corrections, i.e.\ it should be
possible to write an F-theory background whose heterotic dual has the
expected monodromy around $s=0$. This is indeed the case, using for
instance the Shioda-Inose construction described in
\cite{McOrist:2010jw}. Explicitly, we write
\begin{equation}
  y^2 = x^3 + f_\rho xz^4 + g_\rho z^6
\end{equation}
for the $\rho$-fibration, and analogously with $\rho\to\tau$ for the
$\tau$-fibration. The F-theory dual for such a system is then given by
\begin{equation}
  \label{eq:factorisedmap}
  a = -3 f_{\tau} f_{\rho} \, ,\quad b = b_{\tau} b_{\rho} \, , \quad c=
  -\frac{27}{2}g_{\tau}g_{\rho} \, ,\quad d = d_{\tau} d_{\rho} \, ,
\end{equation}
with
\begin{equation}
\Delta_{\tau} = 4 b_{\tau}d_{\tau} \, , \quad \Delta_{\rho} =
4b_{\rho} d_{\rho} \, .
\end{equation}
Whenever $f_\tau$, $g_\tau$, $f_\rho$, $g_\rho$ exist as holomorphic
sections of line bundles of the base (as opposed to the generic
branched form which follows from \eqref{eq:jtau-daleth} and
\eqref{eq:jrho-daleth}) one has that $\daleth$ is a perfect square
globally
\begin{equation}
  \label{eq:daleth-Shioda-Inose}
\daleth = 3^{12} \left[  f_{\tau}^3 g_{\rho}^2-f_{\rho}^3g_{\tau}^2
\right]^2\,.
\end{equation}
Thus, there is no T-duality monodromy around any point in the
base. Equivalently, the condition on the coefficients of the F-theory
fibration for having no T-duality monodromy on the heterotic side is
that the map~\eqref{eq:factorisedmap} holds everywhere in the base for
globally well defined, holomorphic sections 
$f_\tau$, $g_\tau$, $f_\rho$, $g_\rho$. This is clearly not the case for $a$, $b$, $c$, $d$ 
generic sections which will not factor globally on the base.

Coming back to the $I_1$-NS5 system, we take the ansatz
$f_\rho$, $g_\rho$ constant and $f_\tau=-3$, $g_\tau=2+s$. This gives
rise to $\Delta_\tau = 27 s (s+4)$. We have the desired $I_1$
degeneration of the $\tau$ fibration at $s=0$. There is, in addition,
a second $I_1$ degeneration at $s=-4$, which can be included in the
analysis without much trouble. Plugging this ansatz
into~\eqref{eq:factorisedmap} we obtain
\begin{equation}
  a = 9f_\rho\, , \quad b = \Delta_\rho s(s+4)\, , \quad c =
  -\frac{27}{2}(2+s)g_\rho\, , \quad d = \frac{27}{16}
\end{equation}
where we have arbitrarily chosen to locate both small instantons on
the $E_8$ factor
 associated with $b$. By construction, we have that
this ansatz keeps $\rho$ constant, while $\tau$ is non-trivially
fibered. This corresponds to the case of one NS5 on top of the $I_1$
degeneration of the fibration. Note that, as shown in
\cite{McOrist:2010jw}, this configuration of globally constant $\rho$
is the only one accessible in the case of compact heterotic K3 compactifications, if we
insist on having a geometric interpretation of the heterotic
configuration.

\subsection{Moving the NS5 away from the degeneration point}

\begin{figure}
  \centering
  \hfill
  \begin{subfigure}{0.45\textwidth}
    \centering
    \includegraphics[width=\textwidth]{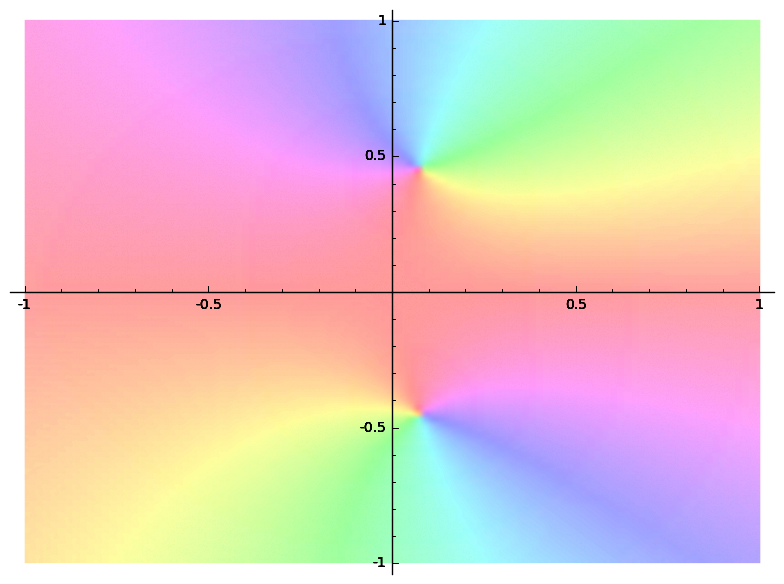}
    \caption{$\epsilon=10^{-1}$.}
    \label{sfig:daleth-large-epsilon}
  \end{subfigure}
  \hfill
  \begin{subfigure}{0.45\textwidth}
    \centering
    \includegraphics[width=\textwidth]{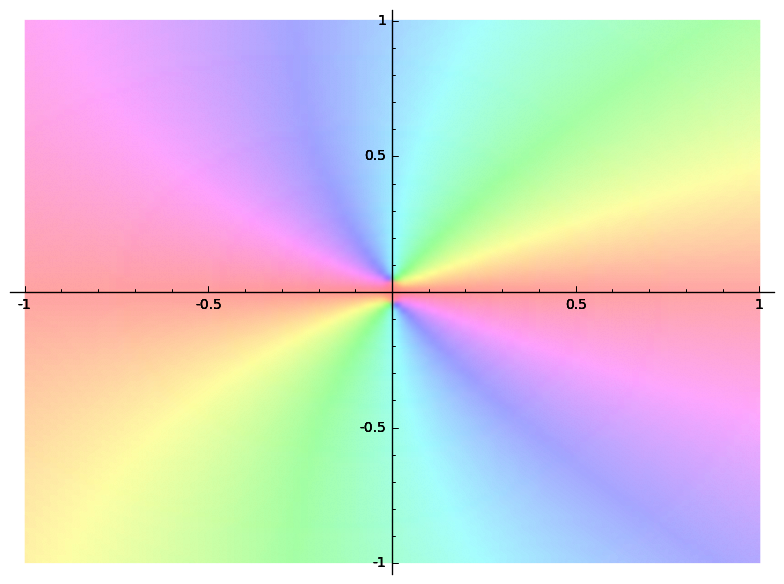}
    \caption{$\epsilon=10^{-3}$.}
    \label{sfig:daleth-small-epsilon}
  \end{subfigure}
  \caption{Plot of $\daleth$ close to $s=0$. The hue of the density plot encodes the phase
    of $\daleth$ while the intensity encodes its modulus. For the two plots we took the ansatz
    \eqref{eq:moved-instanton-ansatz} with $f_\rho=-3$,
    $g_\rho=2+\epsilon$, $\mu=0.5$, and two different values of
    $\epsilon$.  The double zero at $s=0$ in the geometric regime,
    corresponding to the $I_1$ degeneration, splits into two distinct
    single zeroes. These come together as we take $\epsilon\to 0$,
    which corresponds to going to the stable degeneration limit.}
  \label{fig:daleth-split}
\end{figure}

\begin{figure}[t]
  \centering
  \begin{subfigure}{0.45\textwidth}
    \centering
    \includegraphics[width=\textwidth]{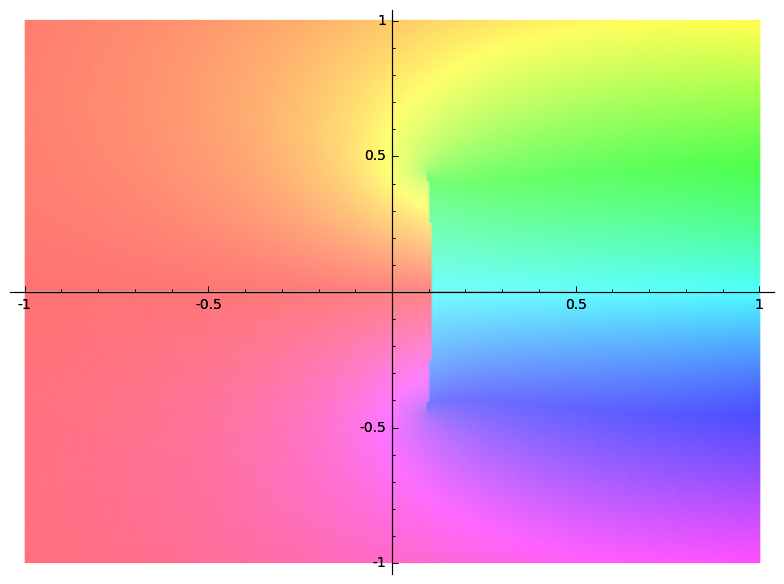}
    \caption{$j(\tau)$.}
  \end{subfigure}
  \hfill
  \begin{subfigure}{0.45\textwidth}
    \centering
    \includegraphics[width=\textwidth]{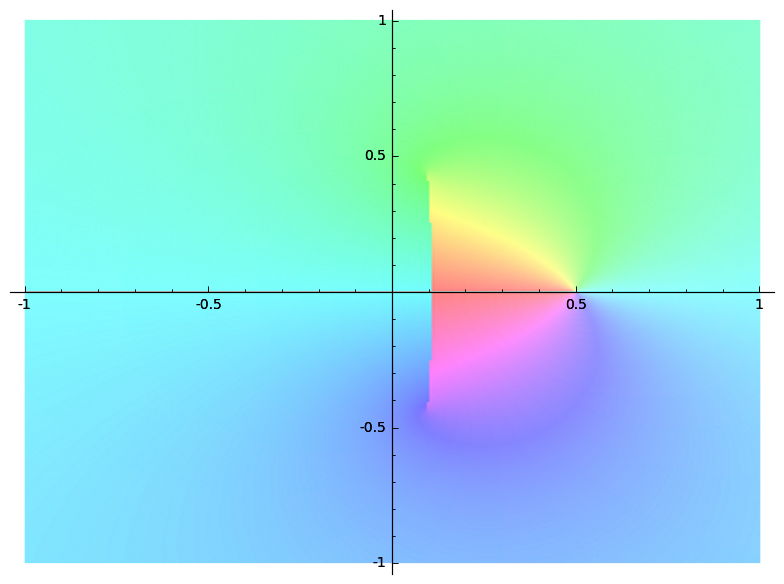}
    \caption{$j(\rho)$.}
  \end{subfigure}
  \caption{Values for $j(\tau)$ and $j(\rho)$ for the parameters of
    Figure~\ref{sfig:daleth-large-epsilon}. The hue encodes the phase
    of the $j$-function, while the intensity encodes its modulus. We
    have labeled the plots according to the classical interpretation
    of the corresponding sheet when $\epsilon\to 0$. The branch cut
    connecting the two components of the $I_1$ degeneration exchanges
    $j(\tau)$ and $j(\rho)$.}
  \label{fig:splitting-plot}
\end{figure}

At this point, a natural question is how to move the NS5 brane away
from the $I_1$ degeneration. We will answer this question in F-theory,
and then dualize back to the heterotic side with the duality map
\eqref{eq:Cardoso-map}. We will see that a geometric interpretation is
lost, and the $I_1$ degeneration splits into its non-geometric components. 
In order to see this quantum splitting, we just need to choose generic
coefficients for the F-theory K3 fibration. For such a generic choice it will no longer be possible to globally factorize
the K3 fibration coefficients such that the
ansatz~(\ref{eq:factorisedmap}) applies with  $f_\rho$, $g_\rho$, $f_\tau$, $g_\tau$ holomorphic sections. We can do this in many ways. Let us
choose for instance to move the small instanton at $s=0$ to $s=\mu$,
while keeping the $a$ and $c$ coefficients unmodified. That is, we
choose
\begin{equation}
  \label{eq:moved-instanton-ansatz}
  a = 9f_\rho\, , \quad b = \Delta_\rho (s-\mu)(s+4)\, , \quad c =
  -\frac{27}{2}(2+s)g_\rho\, , \quad d = \frac{27}{16}\, .
\end{equation}
With this ansatz we have that $\daleth$ no longer factorizes and, therefore, we
obtain non-trivial branch cuts. The discriminant $\daleth$ is in this
case a quartic polynomial, so the expression for the roots is rather
cumbersome and not very illuminating. Instead of writing it here, we
choose to show a plot of $\daleth$ in the neighborhood of $s=0$.
Figure~\ref{fig:daleth-split} shows the splitting of the double
root as we turn on $\mu$. The splitting of the double root into
single roots gives rise to a branch cut in the solutions for $j(\tau)$
and $j(\rho)$, which we display in Figure~\ref{fig:splitting-plot}. 

\subsubsection*{Monodromy of the T-fect}

 The  two non-geometric defects, into which the $I_1$ singularity splits, are
identical up to a choice of duality frame. We denote them (up to duality) by T5. One can easily argue that around a T5
defect one has a monodromy given by
$\sigma\colon(\rho,\tau)\to(\tau, \rho)$, up to a global $O(2,2;\bZ)$
duality transformation. For example, from
\eqref{eq:daleth}-\eqref{eq:jtau-daleth} we notice that the values of
$j(\rho)$ and $j(\tau)$ are identical on top of the T5 and continuously tunable
\begin{equation}
  j(\tau)|_{\daleth=0} = j(\rho)|_{\daleth=0} = -8\sqrt{-\frac{12^3a^3}{bd}}\,.
\end{equation}
Therefore, the square of the monodromy, which leaves $j(\tau)$ and $j(\rho)$
invariant, cannot act non-trivially on either $\tau$ and $\rho$, as
this would require that $\tau$ and $\rho$ are fixed to values
invariant under the corresponding $SL(2,\bZ)$ monodromies, and this is
a discrete set of possibilities. Alternatively, note that the torus
fibers at $\daleth=0$ are generically smooth, and there is no
non-trivial $SL(2,\bZ)$ monodromy compatible with having a smooth
fiber at the origin. This implies that the monodromy action on
$(\tau,\rho)$ is of the form
\begin{equation}
  (\tau,\rho) \to (g\rho, g^{-1}\tau)
\end{equation}
for some $g\in SL(2,\bZ)$. But this is simply a conjugation of
$\sigma$ by the $O(2,2;\bZ)$ element associated with $g$.

\begin{figure}[t]
  \centering
  \begin{subfigure}{0.48\textwidth}
    \centering
    \includegraphics[width=\textwidth]{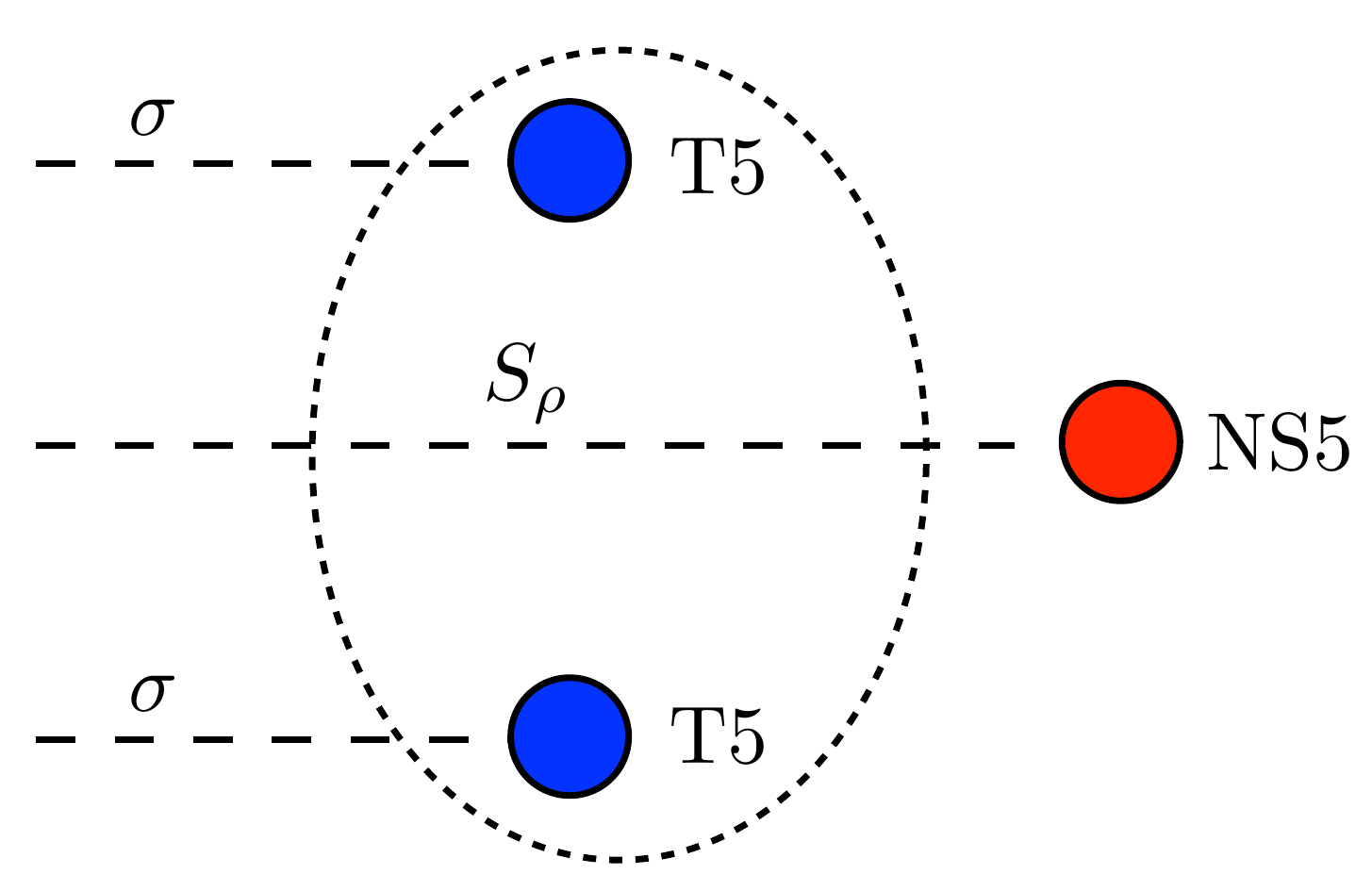}
    \caption{Symmetric presentation.}
    \label{sfig:monodromy}
  \end{subfigure}
  \hfill
  \begin{subfigure}{0.48\textwidth}
    \centering
    \includegraphics[width=\textwidth]{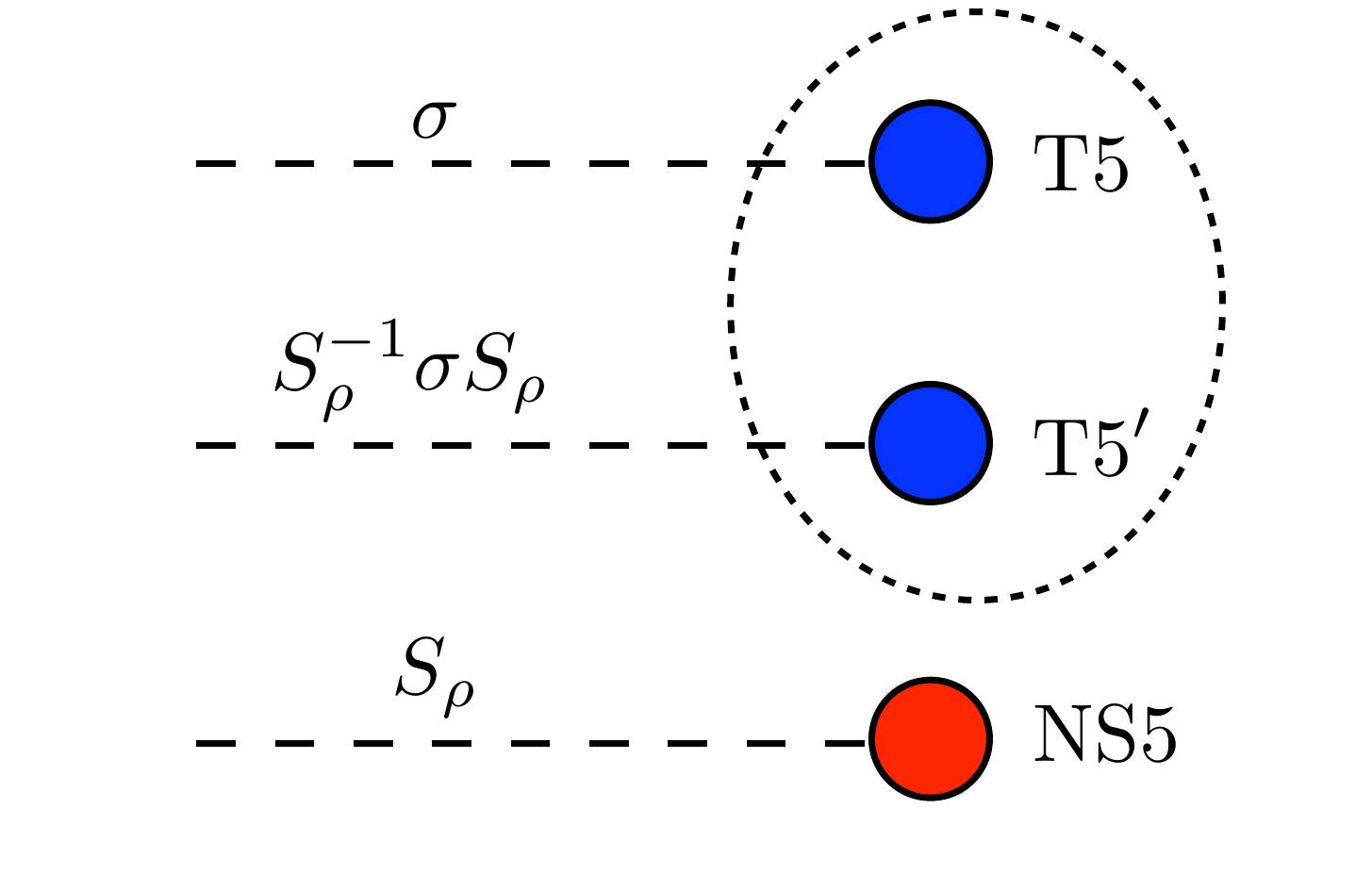}
    \caption{Decoupled presentation.}
    \label{sfig:monodromyNoIntersection}
  \end{subfigure}
  \caption{\subref{sfig:monodromy} Quantum behavior of the $I_1$
    fibration in a heterotic elliptic fibration. The inside of the
    dotted ellipse becomes in the stable degeneration limit a
    geometric $I_1$ degeneration of the elliptic fiber. We have added
    a NS5 brane for convenience, and chosen a convenient description
    for the location of the branch cuts. The monodromy indicated is
    obtained by crossing the branch cut counter-clockwise. Here
    $\sigma\colon (\tau, \rho)\to (\rho, \tau)$,
    $S_\tau\colon(\tau,\rho)\to (\tau+1,\rho)$ and similarly
    $S_\rho\colon(\tau,\rho)\to (\tau, \rho+1)$.
    \subref{sfig:monodromyNoIntersection} An alternate presentation,
    in which the NS5 can be straightforwardly decoupled.}
\end{figure}

While the local monodromy around a T5 brane is of the form $\sigma$,
it must be the case that the monodromy around the two T5 components of
the $I_1$ geometric degeneration is
\begin{equation}
  \kappa \colon (\tau, \rho) \to (\tau+1, \rho -1)\, .
\end{equation}
The way this works is depicted schematically in
Figure~\ref{sfig:monodromy}. We have kept the NS5 in the system for
simplicity (we will remove it momentarily). Let us denote by $S_\rho$
the monodromy associated with the NS5 brane:
\begin{equation}
  S_\rho \colon (\tau, \rho) \to (\tau, \rho + 1)
\end{equation}
and similarly $S_\tau\colon (\tau, \rho) \to (\tau+1, \rho)$. We have
chosen the branch cuts in Figure~\ref{sfig:monodromy} in a way that
makes manifest the local equivalence of the two components of the
resolution of the $I_1$ degeneration. It is then clear (by following
the dotted contour in Figure~\ref{sfig:monodromy}) that in this
description the overall monodromy around the $I_1$ degeneration is
$\kappa=\sigma S_\rho \sigma S_\rho^{-1}=S_\tau S_\rho^{-1}$, as expected.

Let us state the result in the absence of NS5s. We can do so by
starting with Figure~\ref{sfig:monodromy}, and pulling the bottom T5
thought the NS5 branch cut. We obtain in this way a T5 brane in a
different frame, which in Figure~\ref{sfig:monodromyNoIntersection} we
denote by T$5'$. By consistency, the monodromy around the $I_1$
degeneration must stay constant, so we must have that the monodromy of
the T5$'$ brane is given by
\begin{equation}
  M_{\mathrm{T5}'} = S_\rho^{-1}
  \sigma S_\rho= S_\tau S_\rho^{-1}\sigma= \kappa\sigma  \, .
\end{equation}
Now the NS5 brane can be taken away, leaving us with a
consistent description of the $I_1$ degeneration.

Let us remark that the above discussion was done in a convenient
duality frame, but there is a global frame ambiguity having to do with
the $SL(2,\bZ)$ frame chosen for $\tau$. Consider for example
what happens at an $E_6$ degeneration. It splits into 8 $I_1$
singularities which are not all mutually local. They can for instance
be chosen to be of type $(5\times A,B,2\times C)$, in the notation of
\cite{Gaberdiel:1997ud}. Picking a frame in which the monodromy of one
chosen $I_1$ degeneration is $S_\tau$, the monodromy of any other
$I_1$ degeneration will be $(\tau,\rho)\to (S'(\tau), \rho)$, with
$S'=gS_\tau g^{-1}$ and $g\in SL(2,\bZ)$. This can be achieved in the
description above by conjugating every element by the appropriate
element $\hat{g}\in O(2,2;\bZ)$, given by the lift of the $SL(2,\bZ)$
element $g$.

\subsubsection*{No dynamics from the T-fect}

We have seen that the monodromy around each T5 is of the form
$\rho\leftrightarrow\tau$. Thus, on the defect we should have that
$\rho=\tau$. It is well known that the heterotic string compactified on a
$T^2$ develops an enhanced $SU(2)$ gauge group in eight dimensions if the $\tau$ and $\rho$ moduli of the torus are equal. From this, one may erroneously conclude that there is an
$SU(2)$ gauge group on a T5. This
would lead to a failure of anomaly cancellation for compact K3s, and
would also contradict the analysis of \cite{Witten:1999fq}, where it
was found that the heterotic string on ADE degenerations with trivial
background gauge field did not develop any enhanced symmetry.

The resolution of this puzzle is the following. As pointed out in
\cite{Font:2016odl} already, while it is true that on the F-theory
dual the K3 fiber on top of the T5 develops an $A_1$ singularity, the
total space of the fibration of this K3 over $\bC$ is smooth, and thus
there are no new light degrees of freedom associated with the T5.

We can argue this as follows. The elliptic fibration defined
by~\eqref{eq:K3} has a discriminant given by
\begin{equation}
  \Delta = u^{10} v^{10} (27 d^{2} u^{4} + 4 a^{3} u^{2} + 54 c d
  u^{3} + 27 c^{2} u^{2} + 54 b d u^{2} + 54 b c u + 27 b^{2})\, .
\end{equation}
The order 10 roots at $u=0$ and $v=0$ come, as expected, from the two
$E_8$ factors. The resulting discriminant locus
$\Delta'=\Delta/(u^{10}v^{10})$ encodes the position of the remaining
four degenerations of the elliptic fiber, each locally of $I_1$
type. Thus generically the K3 will be smooth away from the $E_8$ points,
but we will have extra singularities whenever some of these four
remaining roots coincide in the $[u: v]$ plane. This occurs whenever
the discriminant of $\Delta'$ as a function of $u$ vanishes,
i.e.\ 
\begin{equation}
  d^{2} \cdot b^{2} \cdot a^{6} \cdot \left(12^3 a^3 bd + \left[4a^3 +
    27(c^2 - 4bd)\right]^2\right) = 0\, .
\end{equation}
We recognize the last factor as $\daleth$ and, thus, we see that
$\daleth=0$ is associated with extra singularities of the K3
fiber. This is as expected, since at $j(\rho)=j(\tau)$ the heterotic
string on $T^2$ has a symmetry enhancement from
$E_8\times E_8\times U(1)^2$ to
$E_8\times E_8 \times SU(2) \times U(1)$. One can show
\cite{LopesCardoso:1996hq} that the singularity in the K3 is of $I_2$
type, which has a local model given by
\begin{equation}
  \label{eq:A1-model}
  \sum_{i=1}^3 z_i^2 = 0
\end{equation}
in appropriate local variables (given by linear functions of $(x, y, u)$
in an affine patch $v=z=1$); we omit the explicit expressions for the
redefinitions since they are not particularly illuminating.

We now take $a$, $b$, $c$, $d$ to be functions of a complex coordinate $\bC$,
and impose that $\daleth$ vanishes to first order in $s$. In terms of
the local model~\eqref{eq:A1-model} this fibration structure is
manifested as a base-dependent complex deformation of the $A_1$
singularity, which in our setup is given by the local form
\begin{equation}
  \sum_{i=1}^4 z_i^2 = \delta
\end{equation}
with $z_i$ given by linear redefinitions of $(x,y,u,s)$, and
$\delta\neq 0$. We recognize this space as the deformed conifold,
which is indeed smooth. So we see that while the K3 fiber is singular
above $s=0$, the total space is smooth, as we claimed above.

\subsubsection*{Stable degeneration limit}

The classical picture of the heterotic string is obtained when the
elliptic $T^2$ is of infinite size.\footnote{We should also require
  that the 10d heterotic string coupling is small, which dualizes to
  the requirement that the $\bP^1$ base of the dual F-theory K3 is
  small \cite{Vafa:1996xn}. Since we will be discussing only the
  complex structure of the K3 we will ignore this requirement in what
  follows.}  In terms of the $j$-functions, we send
$j(\rho)\to \infty$ while keeping $j(\tau)$ generically
finite. Equivalently, we want $\kappa(\rho)\to -1$ keeping
$\kappa(\tau)$
generic.\footnote{\label{footnote:stable-degeneration}Using~\eqref{eq:kappa-fg}
  and \eqref{eq:kappa-Cardoso}, we see that in the limit, and up
  to rescalings, $(f_\tau,g_\tau)=(a,c)$ \cite{Morrison:1996na}.}
This can be achieved by a scaling limit described in
\cite{Morrison:1996pp}. A convenient parameterization of this scaling
limit is to keep $a,c$ constant, while writing $bd=\epsilon b'd'$ and
sending $\epsilon\to 0$. This limit was studied above in connection
with the location of the NS5 branes, and it indeed gives
$j(\rho)\to \infty$ keeping $j(\tau)$ finite. To leading order in
$\epsilon$, we have that $\daleth$ is a perfect square, as
in~\eqref{eq:daleth-NS5}. Since $\daleth$ is a perfect square we have
a well defined geometric interpretation for the background, so we can
globally apply the ansatz~\eqref{eq:factorisedmap}, and we end up with
the factorized form~\eqref{eq:daleth-Shioda-Inose}. Up to an
irrelevant rescaling, for an elliptic curve with $j\to \infty$ one
can choose $(f,g)=(-2,3)$. Plugging this
into~\eqref{eq:daleth-Shioda-Inose}, we read that in the stable
degeneration limit
\begin{equation}
  \daleth = 3^{12}\Delta_{\tau}^2\, .
\end{equation}
We observe that in the stable degeneration limit all of the double
roots of $\daleth$ pair up with roots of the $\tau$ fibration. In
other words, this supports the idea that one should think of the
T-fects as components of the degenerations of the elliptic $\tau$-fibration,
that get resolved by quantum effects.

\subsection{Genericity of the splitting}

\label{sec:genericity}

In the discussion above we have not quite proven that the splitting
always happens unless there is an NS5 brane on top of the $I_1$. This
can be argued in various ways.

First, let us show this directly in a simple example. We choose to
study a case without NS5 branes, given by taking
\begin{equation}
  a = a_0, \quad b = b_0, \quad c = s, \quad d = 1
\end{equation}
in~\eqref{eq:K3}. Both $a_0$ and $b_0$ are arbitrary constants,
parameterizing the moduli space. This is the most general form with
these degrees in $s$, up to coordinate redefinitions, except that we
have chosen to retain $b_0$ general. It can be set to one by a
coordinate redefinition, but this obscures the behavior in the stable
degeneration limit, which in this parameterization is simply
$b_0\to 0$, keeping $a_0$ finite but arbitrary.\footnote{In fact $a_0$
  will parameterize the complex structure of the $T^2$ in the infinite
  volume limit, see
  footnote~\ref{footnote:stable-degeneration}. From~\eqref{eq:kappa-Cardoso}
  we have that ${a_0^3}/{s^2}={27\kappa(\tau)}/{4}$.} We are
interested in the values of $a_0$, $b_0$ for which $\daleth$ has a double
root somewhere in the $s$ plane. Computing the discriminant of
$\daleth$ with respect to $s$ we obtain
\begin{equation}
  \Delta_s(\daleth) \propto b_{0}^{2} \cdot a_{0}^{6} \cdot (a_{0}^{3} + 27
  b_{0})^{2} \, .
\end{equation}
We recognize the $b_0=0$ component as the stable degeneration limit,
as discussed above, where $\daleth$ becomes a perfect square, and we
have two $I_1$ degenerations at $s=\pm\sqrt{-\frac{4a_0^3}{27}}$. (As
we will see momentarily, these $I_1$ singularities are mutually
non-local.)

We also have solutions away from the stable degeneration limit, at
$a_0=0$ and $a_0^3=-27b_0$. Note that since $a_0$ parameterizes the
complex structure of the elliptic curve it is in general arbitrary in
the stable degeneration limit, while the previous families of
solutions would go to $j(\tau(s))=0$ in this 
limit, so that everywhere $\tau=\exp(2\pi i/3)$ up to $SL(2,\bZ)$ conjugation. This
shows that generically the $I_1$ singularity splits as we go away from
the stable degeneration limit.

The special loci with $a_0=0$ correspond in the stable degeneration
limit to having a Kodaira singularity of type II, since
$\deg(f,g,\Delta)=(\infty, 1, 2)$. In this case $\tau$ is constant,
and there is no obvious reason from the supergravity viewpoint why the
singularity should split. And indeed, as we see, it can be taken away
from the stable degeneration limit without inducing non-geometry.

A less hands-on argument goes as follows. Assume that there was a
way of moving the NS5 away from the $I_1$ degeneration without losing
the geometric picture (that is, keeping a globally well defined
separation between $\tau$ and $\rho$). Since the total monodromy of
the system in $\rho$ vanishes, and $\rho\to\rho+1$ around the NS5, we
necessarily have a $\rho\to\rho-1$ monodromy around the $I_1$
degeneration. Supersymmetry requires the fibration to be holomorphic
on the base coordinates, so the putative behavior of the elliptic
fiber around the $I_1$ degeneration should belong to the Kodaira
classification. But there is no degeneration with that monodromy, so
we conclude that no holomorphic fibration with the right properties
exist.

We can also consider what happens when we compactify the base to a
$\bP^1$, so we end up with a compact K3 on the heterotic side and an elliptically fibered 
threefold with base a Hirzebruch surface on the F-theory side. As
discussed in \cite{McOrist:2010jw}, imposing that the heterotic side
has a globally well defined geometric structure fixes rather
dramatically the choice of K3s: one finds that $f_\rho$ and $g_\rho$ should
be holomorphic sections of the trivial bundle over the base $\bP^1$,
which forces them to be constants. In other words, the $B$ field is
constant, which implies that we have local cancellation of the Bianchi
identity, and one NS5 brane on top of every $I_1$ degeneration. This is in
perfect agreement with the local behavior we have discussed.

\section{F-theory interpretation of the $SO(32)$ splitting}
\label{sec:SO(32)}

We now want to discuss the analogous splitting behavior in the context
of the $SO(32)$ heterotic string. As we will see, from the heterotic
viewpoint the conclusions will be identical to those we found in the
$E_8\times E_8$ case. Compactifications of the $SO(32)$ heterotic
string preserving the full $SO(32)$ gauge group dualize to F-theory
backgrounds with simple IIB weak coupling limits. Therefore, our main interest in this regime
will be to understand  the relation between the splitting
we found and the well known splitting of the O7$^-$ plane in IIB. As
we will see, both effects are related.

We will need two ingredients:
\begin{enumerate}
  \renewcommand\labelenumi{\bfseries\theenumi.}
\item\label{item:so32} A map from the $SO(32)$ string to F-theory away from the stable
  degeneration limit. We want in particular to have some way of
  determining $j(\tau)$, $j(\rho)$ for the heterotic $T^2$ in terms of
  the K3 data, analogous to the expressions in
  \cite{LopesCardoso:1996hq}.
\item A notion of stable degeneration limit, so we can make
  statements about quantum effects on a classical background.
\end{enumerate}
The map in point {\bf\ref{item:so32}} is provided in \cite{McOrist:2010jw}, for
example. We have that for an $SO(32)$ heterotic $T^2$ compactification
described by $f_\tau$, $g_\tau$ and $f_\rho$, $g_\rho$ we have a dual
K3 given by
\begin{equation}
  \label{eq:Weierstrass-SO(32)}
  y^2 = x^3 + (u^3v + auv^3 + c v^4)x^2z^2 + bd\, v^8 xz^4
\end{equation}
with
\begin{equation}
  a = -3f_\tau f_\rho \quad ; \quad c = -\frac{27}{2} g_\tau g_\rho
  \quad ; \quad bd = \frac{1}{16} (4f_\tau^3 + 27 g_\tau^2) (4f_\rho^3
  + 27 g_\rho^2)\, .
\end{equation}
We have denoted by $[u\colon v]$ the coordinates on the base $\bP^1$
of the elliptically fibered K3. This K3 is defined by the
hypersurface~\eqref{eq:Weierstrass-SO(32)} on the ambient toric space
\begin{equation}
  \begin{array}{c|ccccc}
    & u & v & x & y & z \\
    \hline
    \bC^* & 1 & 1 & 4 & 6 & 0 \\
    \bC^* & 0 & 0 & 2 & 3 & 1
  \end{array}\,.
\end{equation}
We can now easily check that with these expressions
\begin{subequations}\label{eq:duality-map_SO32}
  \begin{align}
    j(\tau)j(\rho) & = -12^6 \frac{a^3}{27bd}\,,\\
    \lambda(\tau)\lambda(\rho) & = 12^6 \frac{c^2}{4bd}\,,
  \end{align}
\end{subequations}
which are identical to~\eqref{eq:Cardoso-map}. Like in the case of the
$E_8 \times E_8$ heterotic string, \eqref{eq:duality-map_SO32} is the
duality map between the $SO(32)$ heterotic string on $T^2$ and
F-theory on the elliptic K3 \eqref{eq:Weierstrass-SO(32)} that we will
actually use to obtain a six dimensional duality via an adiabatic
fibration over a common base.
In order to do this we fiber the K3 hypersurface~\eqref{eq:Weierstrass-SO(32)}
over a complex base. Let us study the compact case first.
We can obtain it from fibering the K3 in~\eqref{eq:Weierstrass-SO(32)} over a $\bP^1$ which we parameterize
by $[s:t]$. The most general  toric ambient space in this setup is given by (up
to redefinitions)
\begin{equation}
  \label{eq:SO(32)-ambient}
  \begin{array}{c|ccccccc}
    & s & t & u & v & x & y & z\\
    \hline
    \bC^*_1 & 1 & 1 & 0 & \alpha & \beta & \gamma & 0\\
    \bC^*_2 & 0 & 0 & 1 & 1 & 4 & 6 & 0 \\
    \bC^*_3 & 0 & 0 & 0 & 0 & 2 & 3 & 1
  \end{array}
\end{equation}
Homogeneity of~\eqref{eq:Weierstrass-SO(32)} requires
$3\beta=2\gamma$, and $\alpha=\beta$. Demanding that the hypersurface is a Calabi-Yau manifold enforces
\begin{equation}
  3\beta = 2 + \alpha+\beta+\gamma\, .
\end{equation}
From here we can read off that $\alpha=\beta=-4$ and  $\gamma=-6$. This
tells us that the total space is an elliptic fibration over an $\bF_4$
base, as expected \cite{Morrison:1996na}. Notice that we have
assumed that the coefficient of the $u^3vx^2z^2$ term
in~\eqref{eq:Weierstrass-SO(32)} is identically one. This is not
necessary, and dropping this assumption leads to the ``hidden
obstructors'' of \cite{Aspinwall:1996vc}. We will ignore this
possibility for simplicity.
Similarly we find
\begin{equation}
  \label{eq:SO(32)-degrees}
  \deg(a) = 8 \quad ; \quad \deg(c) = 12 \quad ; \quad \deg(bd) = 24
\end{equation}
viewed as homogeneous polynomials in $(s,t)$. Since the form of
the duality map is identical in the parameterization we have chosen,
we can simply read off the expression for the $SO(32)$ $\daleth$
from~\eqref{eq:daleth}. Furthermore, we have that
$\deg(\daleth)=48$, as in the $E_8\times E_8$ case. 

\subsection{The stable degeneration limit}

We want to take the limit $\rho\to i\infty$, which implies
$j(\rho)\to \infty$. We can do this, as in the $E_8\times E_8$ case,
by sending $a^3/bd$ and  $c^2/bd$ to infinity while keeping
$a^3/c^2\approx \frac{27}{4}\kappa(\tau)$ arbitrary. A simple way to
achieve this is to let  $b d$ vanish. To leading order~\eqref{eq:Weierstrass-SO(32)} becomes in this limit
\begin{equation}
  \label{eq:SO(32)-K3-stable}
  y^2 = x^2 (x + Q)
\end{equation}
with $Q=(u^3v + auv^3 +
cv^4)z^2$. 
Since the right hand side of~\eqref{eq:SO(32)-K3-stable} has a double
root at $x=0$, the discriminant of~\eqref{eq:SO(32)-K3-stable} with
respect to $x$ vanishes identically. On the other hand, going to the
standard Weierstra\ss{} form by defining $x=x'-\frac{1}{3}Q$ we find
that $f$ of the elliptic fibration is given by $-\frac{1}{3}Q^2$, and
it is in particular generically nonvanishing away from $Q=0$. Thus we
learn that in the stable degeneration limit the F-theory elliptic
fibration has $j(\tau_{\text{IIB}})\to \infty$ almost everywhere. In
other words, the stable degeneration limit on the heterotic side gives
the weak coupling limit on the F-theory side.

This is not very surprising: the stable degeneration limit suppresses
the world-sheet instanton corrections wrapping the heterotic
$T^2$. Under S-duality these become D1-instantons wrapping the type I
$T^2$, and then two T-dualities on this $T^2$ take us to the IIB side,
with the instantons now being D$(-1)$ branes. But the action of these
instantons is suppressed only by the axio-dilaton $C_0+i/g_s$. Hence, we indeed
expect such a behavior: stable degeneration mapping
to weak coupling.

As in the $E_8\times E_8$ case, in the stable degeneration limit we
have that $f_\tau$ and $g_\tau$ are, up to rescalings, equal to $a$ and $c$, respectively. This implies
that $\daleth$ is proportional to $\Delta_\tau^2$, i.e.\ as in the $E_8\times E_8$ case
$\daleth$ only exhibits double zeroes in the stable degeneration limit. These double zeroes are
coinciding with the positions of the (geometric) complex structure degeneration points of the
heterotic $T^2$ fiber. Put differently, assuming
that there are no small instantons on top of the singularity, the geometric
degenerations only split once we include worldsheet instantons.

In order to have a full understanding of the classical picture, we
also want to understand how to read off the position of the small
instantons in the stable degeneration limit. By an argument entirely
analogous to the one given above for $E_8\times E_8$, we find that
these are located where $bd$ vanishes. As a small sanity check, we
notice from the degrees~\eqref{eq:SO(32)-degrees} that in the compact case we do
indeed have 24 instantons. These are required to  cancel the tadpole in the
compact case.

\subsection{Dualizing the $I_1$ degeneration}

After these preliminaries, let us consider the $SO(32)$ heterotic
background with an $I_1$ degeneration. We wish to consider its F-theory
dual in the stable degeneration limit, and reinterpret the answer in
(weakly coupled) IIB language.

According to the discussion above, an $I_1$ singularity on the $SO(32)$
heterotic side 
can be described by the dual K3 fibration:
\begin{equation}
  \label{eq:I_1-SO(32)-example}
  y^2 = x^3 + (u^3v -3uv^3 + 2(s+1)v^4)x^2z^2 + \epsilon^2 v^8
  xz^4\, .
\end{equation}
We have chosen to locate the $I_1$ degeneration at $s=0$ (there is
another $I_1$ singularity at $s=-2$ in the stable degeneration limit),
gone to the patch $t=1$, and parameterized the stable degeneration
limit by $\epsilon\to 0$. We take $\epsilon^2$ instead of $\epsilon$
in~\eqref{eq:I_1-SO(32)-example} in order to simplify some later
expressions. The discriminant of this elliptic fibration is given by
\begin{equation}
  \label{eq:I_1-SO(32)-Delta}
  \Delta_x = \epsilon^4 v^{18} (4v^6 \epsilon^2 - \hat Q^2)= \epsilon^4 v^{18}  (2v^3\epsilon - \hat Q)(2v^3 \epsilon + \hat Q)
\end{equation}
with $\hat Q=u^3 -3uv^2 + 2(s+1)v^3$. We recognize the $v^{18}$
term as the position of the $SO(32)$ stack: 16 mobile D7 branes on top
of an O7$^-$. From an eight dimensional perspective, the
$(2v^3 \epsilon \pm \hat Q)$ terms must be the six components which
form the remaining three O7$^-$ planes in the weak coupling
limit. Note that since $u\cdot v$ is in the Stanley Reisner ideal of
the toric ambient space~\eqref{eq:SO(32)-ambient}, the polynomials
$\hat Q=0$ and $v=0$ do not have a common solution. Therefore,
$(2v^3 \epsilon + \hat Q)$ does not intersect
$(2v^3 \epsilon - \hat Q)$. In some sense the two branes are
parallel to each other.  In the weak coupling limit, i.e.\ $\epsilon\to 0$,
they lie on top of each other and form an O7$^-$ plane wrapping
$\hat{Q}=0$.\footnote{Since the vanishing locus of $\hat Q$ intersects the $\mathbb P^1$ base of the K3 fiber at three
points, these are the remaining three O7$^-$-planes from the eight dimensional perspective.}

Let us look more closely at $\hat Q=0$ in the vicinity of $s=0$.  This
cubic in $u$ and $v$ factors at $s=0$ as
\begin{equation}
  \hat Q|_{s=0} = (u+2v)(u-v)^2\, .
\end{equation}
From this it would seem like we have some exotic physics at $s=0$,
coming from the `coincident' O7$^-$ planes at
$u=v$.
However this is not
the full story in six dimensions, we also have to keep in mind that we
are dealing with a non-trivial fibration over the
$[s:t]$-base. Therefore, the gradient of $\hat Q$ is three dimensional
and at $s=0$ it is given by:
\begin{equation}
d\hat Q|_{s=0} = 3(u - v)(u+v)\,du -6v(u-v)\,dv + v^2(u + v)\,ds\,.
\end{equation}
Since the gradient does not vanish, we conclude that the O7$^-$
worldvolume is smooth at $s=0$ and we expect no extra degrees of
freedom.

\medskip

Let us now discuss what happens as we go away from the stable
degeneration limit. We are interested, in particular, in the behavior
of the 7-branes away from the $SO(32)$ stack. For this purpose
we go to the patch $v=1$, and take the discriminant of $\Delta_x$ as a
function of $u$. We obtain
\begin{equation}
  \Delta_u(\Delta_x|_{v=1}) = (s^2 - \epsilon^2) \cdot ((s + 2)^2
  - \epsilon^2) \cdot \epsilon^{46} \, .
\end{equation}
We find that away from the stable degeneration limit some of the roots
coincide at $s=\pm \epsilon$, and $s=-2\pm\epsilon$. We expect these
to be the quantum components of the $I_1$ singularities at $s=0$ and
$s=-2$ respectively. For concreteness, we focus on the $s=\pm\epsilon$ points
 which are associated with the original $I_1$
degeneration at $s=0$. One easily checks that at both these points the
K3 fiber has a singularity of type $I_2$, associated with the expected
$SU(2)$ symmetry appearing at the self-duality point of T-duality in
the heterotic frame. To study the situation in more detail, we concentrate on
$s=\epsilon$ for which the discriminant \eqref{eq:I_1-SO(32)-Delta} becomes
\begin{equation}
\begin{split}  \Delta_x|_{s=\epsilon,v=1} & = (u + 2) \cdot (u - 1)^{2} \cdot
  \epsilon^{4} \cdot  ( u^{3} - 3 u + 2 +4  \epsilon)\\
  &= \epsilon^{4} \cdot(u + 2) \cdot(u+2+\tfrac49 \epsilon) \cdot ( u - 1)^{2}\cdot\\
  &\phantom{=}\cdot (u-1+2i\sqrt{\tfrac\epsilon3}-\tfrac29\epsilon)\cdot(u-1-2i\sqrt{\tfrac\epsilon3}-\tfrac29\epsilon)+\mathcal O(\epsilon^6)\, .
\end{split} 
\end{equation}
We can interpret this configuration as the slightly split $B,C$
components of a O7$^-$ when we go away from the weak coupling limit.  At $u=-2v$ and $u=-2v-\frac49 \epsilon$ we have the two components of the ``ordinary'' O7$^-$ from before. The two ``coincident'' O7s from above have recombined such
that two mutually local components of the O7$^-$ planes are still
coincident at $u=v$. The remaining two components moved apart to $u=v\pm2i\sqrt{\tfrac\epsilon3}-\tfrac29\epsilon$. As
in the $E_8\times E_8$ case one can easily check that this $I_2$
singularity in the K3 fiber does not lead to singularities in the
threefold, once we take the dependence on $s$ into account.

In the case of the $SO(32)$ heterotic string, we obtain therefore a
direct relation between the exotic splitting into non-geometry and the
well known splitting of the O7$^-$ plane: the dual of the heterotic
$I_1$ degeneration na\"ively has two O7$^-$ planes coinciding on the
K3 fiber (although as explained the O7$^-$ worldvolume is actually
smooth). As we turn on $g_s$ the action of D$(-1)$ instantons splits
these O7$^-$ planes into components, but there is a remnant of the
original behavior in that at two slightly displaced points in the base
two mutually local components of the O7$^-$ planes still
overlap. These points are dual to the locations of the T5 branes.

\section{Relation to previous work}

\label{sec:previous-work}

The mechanism that we have discussed in this paper explains and
illuminates various points in the literature, as we now discuss.

\subsection{Genus-two formulation of $E_8\times E_7$
  compactifications}

As a first point, it was recently pointed out in
\cite{Malmendier:2014uka} that the moduli space of non-geometric
$E_8\times E_7$ heterotic compactifications on K3 agrees with the
moduli space of F-theory backgrounds with base $\bF_{12}$ with
$E_8\times E_7$ singularities on every K3 fiber, which are also known
to describe \emph{geometric} K3 compactifications of the
$E_8\times E_8$ heterotic strings with instanton number $(0,24)$
\cite{Morrison:1996na,Morrison:1996pp}. It was argued in
\cite{Malmendier:2014uka} that this is an effect of going away from
the stable degeneration limit used in
\cite{Morrison:1996na,Morrison:1996pp}. We have seen that this
explanation is mostly correct, with the refinement that the system
away from the stable degeneration limit can sometimes become geometric
for certain high codimension subspaces of moduli space. We have shown
this statement to be true in the particular case of NS5 branes on top
of $I_1$ degenerations. That the two sets of observations can be
related can be made more precise as follows.

Recall \cite{Giveon:1994fu} that the T-duality monodromy
$\rho \leftrightarrow \tau$ can be embedded into the $O(2,2;\bZ)$
duality group as
\begin{equation}
  g = \begin{pmatrix}
    1 & 0 & 0 & 0\\
    0 & 0 & 0 & 1\\
    0 & 0 & 1 & 0\\
    0 & 1 & 0 & 0
  \end{pmatrix}\, .
\end{equation}
This element has $\det(g)=-1$, and thus it belongs to $O(2,2;\bZ)$ but
not $SO(2,2;\bZ)$. The extension to $O(2,3;\bZ)$, if we assume a
trivial action on the Wilson line, is simply
\begin{equation}
  f = \begin{pmatrix}
    g & 0 \\
    0 & 1
  \end{pmatrix}
\end{equation}
which also has $\det(f)=-1$. This is thus an element of $O(2,3;\bZ)$
not in $SO(2,3;\bZ)$. Being more precise, the element does not involve
parity in the torus, so this is in fact an element of $O^+(2,3;\bZ)$
not in $SO^+(2,3;\bZ)$ (here and below we use notation from
\cite{Malmendier:2014uka}). Such elements act with a minus sign on the
modular form $\chi_{35}(\underline{\tau})$. Thus, T-fects associated
with $\rho\leftrightarrow\tau$ exchanges should have
$\chi_{35}(\underline{\tau})=0$ at their core, and indeed as we
pointed out above
$\daleth\propto q =
\chi_{35}^2(\underline{\tau})/\chi_{10}(\underline{\tau})$ for
compactifications preserving $E_8\times E_8$, so the two requirements
are compatible.

\subsection{Absence of bare $I_m$ singularities in the genus two
  formalism}

Another obscure point which we understand now (and was in fact the
motivation for the present work) is that in the classification of
\cite{Font:2016odl} there seemed to be no configuration associated
with heterotic ADE degenerations without NS5 branes on top. For
instance, the simplest $I_{m-n-0}$ singularity for the genus two
fibration was interpreted as $m+n$ NS5 branes on an $I_{\min(m,n)}$
singularity, so one could never entirely remove the NS5 branes. Now we
can explain this observation simply as the statement that such
degenerations without NS5 branes do not appear in heterotic string
compactifications, since they get split into separate non-geometric
components.

Furthermore, as we discussed above, we have that the T-fects resulting from
geometric singularities splitting are elements of $O^+(2,3;\bZ)$ but
not of $SO^+(2,3;\bZ)=\Sp(4;\bZ)$. The classification of
\cite{ogg66,Namikawa:1973yq} only considered monodromies in
$\Sp(4;\bZ)$, and thus the $\rho\leftrightarrow\tau$ T-fects
considered in this paper were not included in \cite{Font:2016odl}.

\subsection{Non-geometric unfreezing of moduli spaces}

A third previously confusing point was the observation in
\cite{McOrist:2010jw} that the moduli space of the heterotic K3
compactifications did not seem to allow to move the NS5 branes away
from the $I_1$ degenerations. As we have seen, the resolution of this
puzzle is that the moduli space with a geometric interpretation is a
subspace of the true moduli space of the heterotic compactification,
which is non-geometric at generic points.

\subsection{A subtlety at low instanton number}

\label{sec:low-instanton-subtlety}

Finally, we would like to highlight a subtlety that becomes apparent
in our analysis, regarding the physics of small instantons at ADE
singularities in the heterotic string. We focus in particular on the
case of the $E_8\times E_8$ heterotic string with $k$ NS5 branes
probing an $I_m$ degeneration. This system (or more precisely, its
$\bC^2/\bZ_m$ limit) was studied in \cite{Aspinwall:1997ye}, with the
conclusion that for $k<2m$ one needs to replace $m$ by
$\left\lfloor\frac{k}{2}\right\rfloor$ in order to obtain the relevant
physics. This result becomes rather puzzling once we embed the
$\bC^2/\bZ_m$ singularity into a $T^2$ fibration, since it breaks
invariance under T-duality, which exchanges $m$ with $k-m$, keeping
$k$ invariant. (We will discuss what happens when $k<m$ below.)
Accordingly, comparison with the results in \cite{Aspinwall:1997ye}
lead the authors of \cite{Font:2016odl} to give an identification of
the physics of the genus two $I_{m-n-0}$ degeneration in terms of
$m+n$ NS5 branes on $I_{\min(m,n)}$ singularities, which breaks the
manifest T-duality invariance of the underlying genus two formalism.

The resolution of the puzzle has to do again with the quantum
splitting of the $I_m$ singularities. It is easy to check that for
$k<2m$ the F-theory background chosen in \cite{Aspinwall:1997ye} in
the stable degeneration limit for representing the physics of $k$
instantons on $\bC^2/\bZ_m$ cannot be continued to finite fiber size
keeping a zero of order $2m$ in $\daleth$, which is a necessary
condition for having a pointlike $\bC^2/\bZ_m$ singularity in the
base. In order to illustrate this point, let us consider a small
modification of the background in \cite{Aspinwall:1997ye}, which has
the same physical properties but is algebraically easier to
deal with:
\begin{equation}
  \label{eq:AM-0}
  a = -3\quad ; \quad b = s^{k}(4+s^m) \quad ; \quad c =
  2+s^m \quad ; \quad d = 0\, ,
\end{equation}
where we used the notation in~\eqref{eq:K3}. If we continue to finite
fiber size by modifying this to
\begin{equation}
  \label{eq:AM-epsilon}
  a = -3\quad ; \quad b = s^{k}(4+s^m) \quad ; \quad c =
  2+s^m \quad ; \quad d = \epsilon
\end{equation}
we obtain
\begin{equation}
  \daleth = 729\, (4 + s^m)\, (4 + s^m - 4\epsilon s^k) \,
  \underbrace{(s^{2m} - 4\epsilon s^{k+m} - 16 \epsilon
    s^k)}_{\text{roots near $s=0$}}\, .
\end{equation}
For $k \geq 2m$ we expect to have no issue with quantum splittings.
In this case $\daleth$ has a zero of order $2m$ at $s=0$ which is
compatible with the presence of an $I_m$ degeneration. A more detailed
computation, using~\eqref{eq:jtau-daleth} and \eqref{eq:jrho-daleth},
shows that the local structure of the heterotic fibration close to
$s=0$ is indeed of the expected form\footnote{Since we are interested
  in the behaviour of the $j$-functions for $|s| \ll \epsilon$, we
  expand~\eqref{eq:jtau-daleth} and \eqref{eq:jrho-daleth} in terms of
  $s$ to obtain the right leading terms.}
\begin{equation}
  j(\tau) \sim \frac{1}{s^m} \quad ; \quad j(\rho) \sim
  \frac{1}{s^{k-m}}\, ,
\end{equation}
i.e.\ $k$ NS5 branes on an $I_m$ degeneration. 
On the other hand, we see that for $k<2m$ $\daleth$ has a zero of order $k$ at $s=0$. More in detail, one can see that close to
$s=0$ we have
\begin{equation}
  j(\tau) \sim \frac{1}{s^{\frac{k}{2}}} \quad ; \quad j(\rho) \sim \frac{1}{s^{\frac{k}{2}}}
\end{equation}
as expected for the system of $k$ NS5 branes on an $I_{\frac{k}{2}}$
singularity, which is consistent with the prescription in
\cite{Aspinwall:1997ye}. For $m \leq k < 2m$ it is thus not correct to
claim that there is an $I_m$ degeneration at the origin. Because for
the ansatz \eqref{eq:AM-epsilon} the singularity splits and leaves us
(effectively\footnote{One can check that the half-integral part of
  $\frac{k}{2}$ for $k$ odd has no effect on the gauge algebra and
  multiplet content on the tensor branch of the SCFT, as obtained from
  the dual F-theory resolution pattern.})  with an
$I_{\left\lfloor\frac{k}{2}\right\rfloor}$ singularity, with
$\left\lfloor\frac{k}{2}\right\rfloor<m$, at the origin.

Alternatively, for $k\geq m$ one can choose to use the Shioda-Inose
ansatz~\eqref{eq:factorisedmap} in constructing the system of $k$ NS5
branes on top of the $I_m$ singularity for finite
$T^2$ volume. That is, we take
\begin{equation}
  \label{eq:SI-epsilon}
  f_{\rho} = -3 \quad ; \quad g_\rho = 2 + \epsilon t^{k-m} \quad ; \quad
  f_\tau = -3 \quad ; \quad g_\tau = 2 + t^{m}
\end{equation}
and
\begin{equation}
  b_\tau = \frac{1}{4}\Delta_\tau \quad ; \quad b_\rho =
  \frac{1}{4}\Delta_\rho \quad ; \quad d_\tau = d_\rho = 1\, ,
\end{equation}
which is an arbitrary choice of the instanton embedding.
In this setting the physics is manifestly invariant under T-duality,
i.e.\ invariant under the $\rho\leftrightarrow\tau$ exchange, and there is no splitting
of the degeneration at the origin. The resulting singularity in
F-theory, and accordingly the resulting SCFT agrees for $k\geq 2m$
with the one obtained from~\eqref{eq:AM-epsilon}, and generically
disagrees for $k<2m$. For instance, for $k=m$ we obtain the same SCFT
as that of $k$ NS5 branes in flat space, with no gauge group, while
the prescription of \cite{Aspinwall:1997ye} gives a non-trivial gauge
group on the tensor branch.

For $k<m$ we cannot write a Shioda-Inose ansatz (since the monodromy
on $\rho$ is negative). In this range it is ill-defined to say in any
case that we have $k$ NS5 branes on $I_m$, since there will be
splitting into non-geometry no matter which ansatz we take. In
continuing to finite fiber volume from $\bC^2/\bZ_m$ we have a genuine
choice: we can either still choose the ansatz~\eqref{eq:AM-epsilon}
(i.e.\ effectively setting $m=\left\lfloor\frac{k}{2}\right\rfloor$),
or choose keeping the degree of $\daleth$ maximal (effectively setting
$m=k$). The second choice leads again to no algebra, agreeing with the
claim in \cite{Hanany:1997gh}.

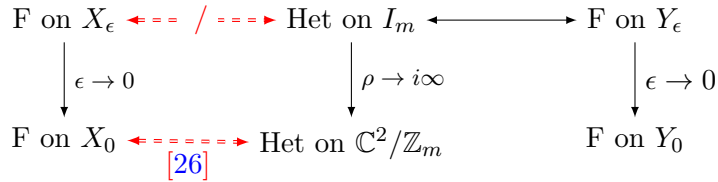
\begin{figure}
  \centering
  \begin{tikzpicture}
    \node (Het-Im) at (0,0) {Het on $I_m$};
    \node[below=of Het-Im] (Het-Zm) {Het on $\bC^2/\bZ_m$};

    \node[left=2cm of Het-Im] (F-EA-eps) {F on $X_\epsilon$};
    \node[below=of F-EA-eps] (F-EA-0) {F on $X_0$};

    \node[right=2cm of Het-Im] (F-SI-eps) {F on $Y_\epsilon$};
    \node[below=of F-SI-eps] (F-SI-0) {F on $Y_0$};

    \draw[-latex] (F-EA-eps) -- (F-EA-0) node[midway, right]
    {\footnotesize$\epsilon\to 0$};

    \draw[-latex] (Het-Im) -- (Het-Zm)
    node[midway, right] {\footnotesize$\rho\to i\infty$};

    \draw[latex-latex] (Het-Im) -- (F-SI-eps);
    \draw[-latex] (F-SI-eps) -- (F-SI-0)
    node[midway, right] {$\epsilon\to 0$};

    \draw[latex-latex, dashed, double, red] (F-EA-eps) -- (Het-Im)
    node[midway, fill=white] {$/$};

    \draw[latex-latex, dashed, double, red] (F-EA-0) -- (Het-Zm)
    node[midway, below] {\cite{Aspinwall:1997ye}};
  \end{tikzpicture}

  \caption{Heterotic/F-theory duality structure for $m \leq k<2m$. By
    $X_\epsilon$ we denote the family of Calabi-Yau
    threefolds~\eqref{eq:AM-epsilon}, while by $Y_\epsilon$ we denote
    the family~\eqref{eq:SI-epsilon}. The struck out dashed arrow on
    the top row indicates that the heterotic dual of F-theory on
    $X_\epsilon$ does not have an $I_m$ singularity at the origin. The
    dashed line on the bottom row is the duality assumed by
    \cite{Aspinwall:1997ye}. The SCFTs associated to F-theory on
    $X_\epsilon$ and $Y_\epsilon$ are generically different.}
  \label{fig:few-instantons-dualities}
\end{figure}

Notice that this freedom in how to move away from the stable
degeneration limit can have rather dramatic effects on the SCFT: in
the cases where compactifying the fiber gives a quantum deformation of
the singularity this translates into a \emph{relevant} deformation of
the SCFT, no matter how large the $T^2$ is, as long as it is
finite. And since in this context we are discussing the SCFT field
theory at the end of the flow, there can be a sharp discontinuity in
what we mean by the SCFT associated with $k$ instantons on $I_m$
depending on whether we are at finite volume or strictly infinite
volume (i.e. $\bC^2/\bZ_m$). This makes the use of the  heterotic/F-theory
duality for determining the physics of $k$ instantons on the $\bC^2/\bZ_m$
singularity and $k<2m$  rather subtle. In this case there are
various families of quantum corrected $T^2$ fibrations which
decompactify to $\bC^2/\bZ_m$, but which have rather different physics
for any finite value of the fiber size. We summarize the situation in
Figure~\ref{fig:few-instantons-dualities}.

\section{Discussion}
\label{sec:discussion}

We have shown that the heterotic string on an elliptically fibered K3
(local or global) with small instantons, dynamically develops a
non-geometric structure whenever there are not enough NS5 branes on
the geometric degenerations to make the contribution of the local
system to the modified Bianchi identity everywhere
non-negative.\footnote{As pointed out in \S\ref{sec:genericity} there
  can be exceptions to this statement whenever the local degeneration
  is compatible with having a constant $\tau$, which is a very
  non-generic situation.} In the case of the compact smooth K3 there
is a very special locus in moduli space where we have exactly one NS5
on top of each $I_1$ degeneration. In this case we recover the
geometric interpretation, but any deformation away from this point
will make the background non-geometric. As we have discussed in
\S\ref{sec:previous-work}, the existence of this non-geometric
splitting resolves a number of puzzles in the literature.

\medskip

While in this paper we have focused on the simple case of small
instantons on K3, it is easy to see that our observation will
generalize to a large class of interesting heterotic backgrounds. For
instance, we could think about starting from a generic configuration
where the NS5 branes are away from the $I_1$ degenerations, and going
onto the Higgs branch. That is, we ``dissolve'' the NS5 branes into a
smooth bundle. By continuity, we learn that generic heterotic bundles
will then also give rise to T-folds, and only over some very specific
conditions a standard geometric interpretation will be possible. (As
mentioned above, this genericity of non-geometry was already observed
in \cite{Malmendier:2014uka} in the case of $E_8\times E_7$
compactifications on K3 with instanton numbers $(0,24)$, which
suffices for making the bundle completely smooth.)  Let us emphasize
that we are not claiming that every smooth bundle may induce
non-geometry, only that this is the generic nature of heterotic
compactifications. Specific choices of the bundle might still be
geometric. A possible candidate to stay geometric would be the
standard embedding, which is known to have the same spectral cover as
the system of point-like instantons on $I_1$ degenerations
\cite{Friedman:1997ih,Bershadsky:1997zv,Aspinwall:1998he,Donagi:2011dv},
and cancels the Bianchi identity pointwise.

Clearly, non-geometric splitting will also be the generic case for
compactifications to fewer than six dimensions, as we can apply the
six-dimensional analysis close to neighborhoods of the discriminant
locus of the elliptic fibration.

It is well known that existing techniques with high probability give
elliptically fibered Calabi-Yaus \cite{Rohsiepe:2005qg,Candelas:2012uu,Johnson:2014xpa,Gray:2014fla,Johnson:2016qar,Anderson:2016cdu}, so our observation
directly applies to the heterotic string on many of the available
backgrounds. It would nevertheless be important to understand if a
similar non-geometric splitting holds when there is no elliptic
fibration structure. We cannot directly extrapolate our results to
this setting, as the heterotic/F-theory duality map is no longer
available.

\acknowledgments

We thank James Gray, Diego Regalado, Sav Sethi, Timo Weigand and Alberto
Zaffaroni for illuminating discussions, and particularly Anamaría Font
for collaboration on related earlier work.

\bibliographystyle{JHEP}
\bibliography{refs}

\end{document}